\documentclass[a4paper,11pt]{article}
\usepackage{pos}
\usepackage{float}
\usepackage{caption}

\title{Advanced stereoscopy applied to CTAO}

\author*[a]{Hana Ali Messaoud}
\author[a]{Tom François}
\author[a]{Thomas Vuillaume}

\affiliation[a]{Laboratoire d’Annecy de Physique des Particule,CNRS\\
  LAPP, 9 chemin de bellevue, 74940 Annecy, France}

\emailAdd{hana.alimessaoud@lapp.in2p3.fr}

\abstract{The Cherenkov Telescope Array Observatory (CTAO) is an international observatory currently under construction, which will consist of two sites (one in the Northern Hemisphere and one in the Southern Hemisphere). It will eventually be the largest and most sensitive ground-based gamma-ray observatory. In the meantime, a small subarray composed of four Large-Sized Telescopes (LSTs) at the Northern site will begin collecting data in the coming year.
In preparation, we present a stereoscopic event reconstruction using graph neural networks (GNNs) to combine information from several telescopes of this subarray. In our previous work, we explored the use of GNNs for the stereoscopic reconstruction of gamma-ray events on simulated data from the Prod5 sample and showed that GNNs provide a better stereoscopic reconstruction. We now compare this approach to the currently foreseen method that analytically combines the output of monoscopic random forests, and explore how GNNs can be used in fusion with the Random forest algorithm in order to provide a more sensitive stereoscopic system.}

\ConferenceLogo{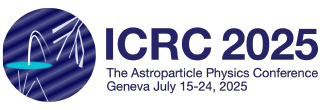}

\FullConference{39th International Cosmic Ray Conference (ICRC2025)\\
 15–24 July 2025\\
Geneva, Switzerland\\}

\begin{document}
\maketitle
\section{Introduction}
CTAO will observe cosmic ray events indirectly by capturing the Cherenkov light produced by their interaction with the atmosphere. 
The event reconstruction is the analysis of the captured images to determine the physical properties (mainly energy, incoming direction and particle type) of the cosmic ray. 
One step of this reconstruction process is the stereoscopy, a method that combines data from several telescopes to reconstruct the energy, direction, and classification of the event.
The standard method as currently implemented in ctapipe  \citep{linhoff2023ctapipe}, the processing pipeline software for CTAO, combines individual telescope reconstructions using weighted averages (for class and energy estimation) and geometric methods for the direction.
This approach is straightforward and has shown consistent, good results \citep{linhoff2023ctapipe}.
In this work, we  propose a novel stereoscopic reconstruction based on Graph Neural Networks (GNNs) for stereoscopic event reconstruction at the Cherenkov Telescope Array Observatory (CTAO). We apply GNNs to data from the prod5 sample, using all telescopes at the CTAO North site, 
We also combine the results from random forests with those from the GNNs to see if this improves the reconstruction performance.
\section{Datasets}
We use Monte Carlo simulations generated with the CORSIKA software \citep{heck2019corsika}, which simulates atmospheric showers induced by high-energy cosmic rays. The telescope response to the signal is then simulated using the simtelarray software \citep{bernlohr2008simulation}. The simulation sample used is prod5b\footnote{https://zenodo.org/records/5499840}.
The data is afterwards analyzed using ctapipe v0.19.0 \citep{linhoff2023ctapipe} up to data level DL2, which includes simulated parameters (ground truth), image parameters, and reconstructed event parameters, used here as the baseline. 
\section{Methods}
\subsection{Random Forests+Hillas}
The stereoscopic reconstruction of events using Random Forest + Hillas method involves two main steps. 
The first step is the monoscopic reconstruction, which is based on observations from a single telescope at a time. It is used to predict the energy and direction of the incident particle using a Random Forest algorithm. 
The direction is reconstructed analytically and geometrically, while energy prediction and classification are performed using a Random Forest (RF) algorithm for each telescope type.
The stereoscopic reconstruction is then performed by weight-averaging the energy and gammaness predictions.
As for the direction, it is reconstructed geometrically by combining the images of the same event observed by several telescopes and the intersection of the major-axis of the ellipses gives the position of the source.
\subsection{Stereograph: Graph Neural Network-based Reconstruction}
Stereograph is a Graph Neural Network based model for stereoscopic event reconstruction at the Cherenkov Telescope Array Observatory (CTAO).
In our approach, each event is represented as a separate graph, where the nodes correspond to the telescopes that observed the same event.
When an atmospheric shower is observed by multiple telescopes, edges are created in the graph to connect the nodes corresponding to the telescopes that have observed the same event. Thus, each unique observation of an atmospheric shower captured by an array of telescopes is then represented by a graph.
Each node has as features a set of parameters extracted from telescope observations, known as Hillas parameters. These parameters describe the geometric characteristics of air shower images, as introduced by Hillas\citep{Hillas1985}. The method assumes that the image produced by a gamma-ray  has an elliptical shape. After cleaning the images, the ellipsoid is afterward characterized by its moments up to second order. In total, each node has 30 features, which are mainly Hillas parameters.
Since our baseline comparison is the [RF+Hillas] method, we used the same features used to train the Random Forest model.
\section{Metrics}
\subsection{AUC/ ROC }
The Receiver Operating Characteristic (ROC) curve is used to evaluate the performance of a classification model. It plots the true positive rate against the false positive rate for different classification thresholds.
The Area Under the Curve (AUC) is defined as the integral of the ROC function. 
The gammaness, representing the probability that an incident particle is a gamma rather than a proton, indicates the confidence level that an event originates from a gamma ray.
For the gamma/proton classification task, the overall performance of the network is given by the area under the ROC curve (AUC). 
The AUC is computed over the entire dataset.
\subsection{Energy resolution}
Energy resolution, represents, for each energy bin, the interval around 0 that contains 68$\%$  of the distribution of the relative prediction error $\frac{\delta{E}}{E}$. It serves as a measure of error, and we aim to keep it as low as possible.
\subsection{Point spread function (PSF)}
Similarly to the energy resolution,  the angular resolution represents, for each energy bin, the interval that contains 68\% of the distribution of the angular separation between predictions and true directions.
\clearpage
\section{Application of Graph Neural Networks to the stereoscopic reconstruction of gamma-ray events}
We applied stereograph to the array composed of 13 telescopes, 4 of which are Large-Sized Telescopes (LST) and 9 of the Medium-Sized Telescopes (MST).\\
After training our model, we computed the IRFs using ctapipe-compute-irfs-tool.\\
We evaluate the classification model using the Receiver Operating Characteristic (ROC) curve, observing that the classification performance improves with a higher AUC across all energy ranges, which shows that the model is effective at distinguishing gamma rays from protons.

\begin{figure}[H]
    \centering
    \includegraphics[width=1\linewidth]{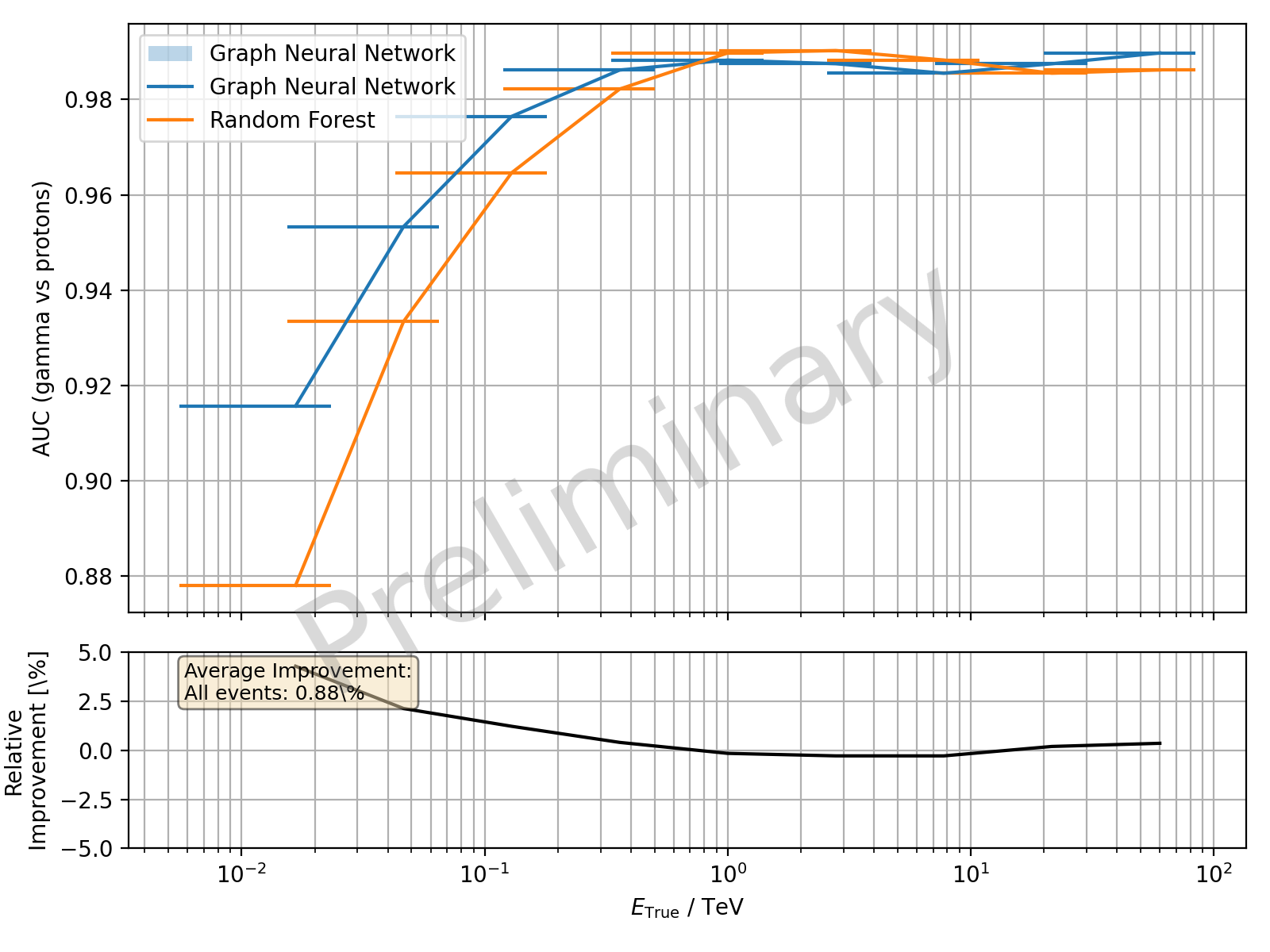}
       \caption{The curve shows AUC values with respect to energy bins. It shows the comparison between the results obtained using graph neural networks and the baseline (RF + Hillas).}
    \label{fig:auc}
\end{figure}

 As shown in Figure~\ref{fig:auc}, the model performs better at both low and high energies, while showing a slightly reduced performance in the intermediate range between 1 TeV and 10 TeV.
We use these results to select the most gamma-like events in each energy bin based on their reconstructed gammaness values, where gammaness denotes the confidence that an event was produced by a gamma ray.
The angular resolution was calculated using geometric reconstruction methods. It was then used to apply a $\theta$-cut. 

\begin{figure}[H]
    \centering
    \includegraphics[width=1\linewidth]{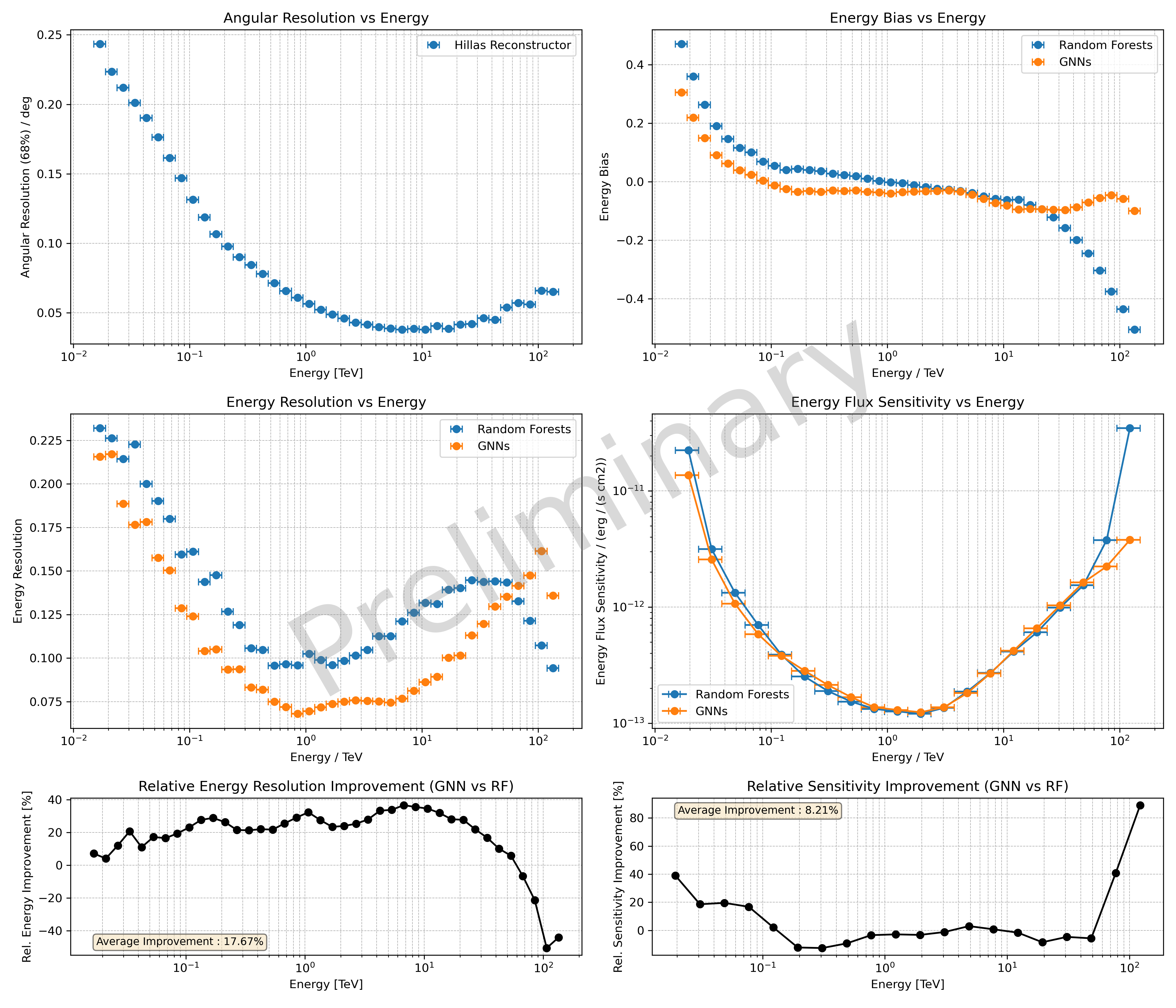}
    \caption{Comparison of IRFs between the standard Random Forest (RF+Hillas) approach and the proposed Graph neural networks based method (GNN+Hillas). 
    The plots show improved energy resolution and reduced bias with the GNN method. Angular resolution is computed geometrically in both methods, and the sensitivity is slightly better.}
    \label{fig:sensitivity}
\end{figure}After optimizing the IRFs with the same tool, we compared the performance of both methods (see Figure~\ref{fig:sensitivity}).
Overall, our approach results in improved energy resolution and reduced bias, as well as a slightly better sensitivity.


\section{Conclusion}
Graphs are very promising for improving the stereoscopic reconstruction of gamma events. As we saw an improved energy resolution and bias, GNNs also performed well in distinguishing gamma rays from hadrons.
Graph-based methods seem like a natural and effective choice for this kind of reconstruction. However, a big challenge with Graph Neural Networks (GNNs) is that they are hard to understand and explain. Even so, GNNs still show great potential for improving gamma event reconstruction.

\end{document}